\documentclass[a4paper,11pt]{article}
\usepackage{pos}
\usepackage{braket}
\usepackage{multirow}

\title{Update on the computation of the quenched $SU(6)$ Yang-Mills lattice spectrum}
\ShortTitle{Update on the computation of the quenched $SU(6)$ YM lattice spectrum}

\author*[a,b]{Andrea Falzetti}
\author[a]{Matteo Lombardi}
\author[a,b]{Mauro Lucio Papinutto}
\author[a,b]{Francesco Scardino}

\affiliation[a]{University of Rome "La Sapienza",\\
  Piazzale Aldo Moro 2, Rome, Italy}

\affiliation[b]{Istituto Nazionale di Fisica Nucleare (INFN)
Sezione di Roma\\
Piazzale Aldo Moro 2, Rome, Italy \\
G. Marconi building}

\emailAdd{andrea.falzetti@uniroma1.it}
\emailAdd{matteo.lombardi@uniroma1.it}
\emailAdd{mauro.papinutto@uniroma1.it}
\emailAdd{francesco.scardino@uniroma1.it}

\abstract{We report on our continued efforts to measure the glueball and meson spectra in SU($N$) Yang-Mills theory and QCD with the aim of extrapolating to the large-$N$ limit. In particular, we document the computation of the low-lying SU($6$) spectrum. We employ a multilevel sampling algorithm to measure glueball correlators to reduce statistical noise in the large-time separation limit. The gluon operator basis is composed of spatial Wilson loop measured at different levels of (APE) smearing, with vanishing momentum selected to maximise the orthonogality of the operators and their overlap with the lowest lying states. We also report on analogous computations for the $J=0,1$ non-singlet meson spectrum with two degenerate quark flavors.}

\FullConference{The 42nd International Symposium on Lattice Field Theory (LATTICE2025)\\
2-8 November 2025\\
Tata Institute of Fundamental Research, Mumbai, India\\}

\begin{document}
\maketitle

\section{Introduction}
Quantum Chromodynamics (QCD) is the theory of the strong interactions. It is a non-Abelian gauge theory with gauge group $SU(N)$ coupled to fermionic matter, with $N=3$ in the real world. Even its pure-gauge sector, namely $SU(N)$ Yang--Mills theory, has a highly non-trivial dynamics for $N>1$~\cite{YM_original}. Its spectrum contains the so-called glueball states, created by gauge-invariant composite operators. The existence of glueballs is one of the most important theoretical predictions of QCD~\cite{fritzsch2002currentalgebraquarkselse}, and also one of the most difficult to confirm experimentally. Indeed, searches for glueball states have been ongoing for decades~\cite{Ochs_2013,Crede_2009}; nevertheless, the experimental determination of the glueball spectrum remains extremely difficult because of the strong mixing with flavour-singlet meson states predicted by the theory. As a consequence, although several glueball candidates have been proposed, no definitive direct determination of a glueball mass in real-world QCD has yet been achieved with complete confidence.\par
At high energies, asymptotic freedom allows QCD to be successfully studied in perturbation theory, for instance in deep inelastic scattering~\cite{asym_free1,asym_free2}. However, because of confinement, we are presently unable to derive analytically either the physical hadronic spectrum or the full phenomenology of hadronic physics, let alone the exact $\mathcal{S}$-matrix of the theory. In particular, the appearance of the renormalisation-group (RG) invariant mass scale $\Lambda_{\mathrm{QCD}}$, which vanishes at every order in perturbation theory, implies that perturbative methods fail precisely for the physical masses of the theory.
For this reason, the most successful first-principles approach to QCD is its regularisation on a spacetime lattice, followed by a numerical evaluation of the functional integral~\cite{WilsonLattice,Berg1,Berg2}. The glueball spectrum can then be extracted from the large-time behaviour of two-point correlators of purely gluonic operators. This provides a natural first-principles framework for determining the spectrum.\par
A complementary analytical perspective is provided by the large-$N$ expansion introduced by 't Hooft, in which the $SU(3)$ colour gauge group is generalized to $SU(N)$ and observables are expanded in powers of $1/N$ around the $N\to\infty$ limit~\cite{thooft-largeN}. In the double-line representation, the corresponding Feynman diagrams are topologically organised by the genus expansion of punctured Riemann surfaces, each topology being weighted by a factor $N^\chi$, where $\chi=2-2g-n$ is the Euler characteristic~\cite{thooft-largeN,Veneziano1976}. This structure matches that of a string perturbation theory with closed-string coupling $g_s=1/N$~\cite{Veneziano1974}. This correspondence has long been taken as evidence for the existence of a non-perturbative string description of large-$N$ Yang--Mills theory and large-$N$ QCD~\cite{thooft-largeN,Veneziano1976,Aharony2000}. In such a picture, glueballs and mesons arise respectively as closed- and open-string excitations, and in asymptotically free theories the string tension is expected to be set by the RG-invariant scale of the gauge theory~\cite{Bochicchio2018}. In the $N\to\infty$ limit, the theory simplifies considerably, becoming an effective theory of infinitely many weakly interacting mesons and glueballs.\par
The long-term goal of the present study is to test a Semi-Topological String Field Theory (STFT) approach defined along the lines of Ref.~\cite{bochicchio_asymSol}. This framework leads to very rigid predictions for the glueball spectrum in the 't Hooft limit, in particular in terms of exactly linear Regge trajectories~\cite{Bochicchio2018}. Our aim is to push the numerical precision of lattice determinations by combining modern hardware accelerators with multilevel sampling techniques, and to use the resulting lattice data to validate or falsify this scenario. In this contribution, we present measurements of the glueball and meson spectra for the gauge group SU($6$).

\section{Glueball Operator Basis}
\begin{figure}[h!]
    \centering
    \includegraphics[width=0.5\linewidth]{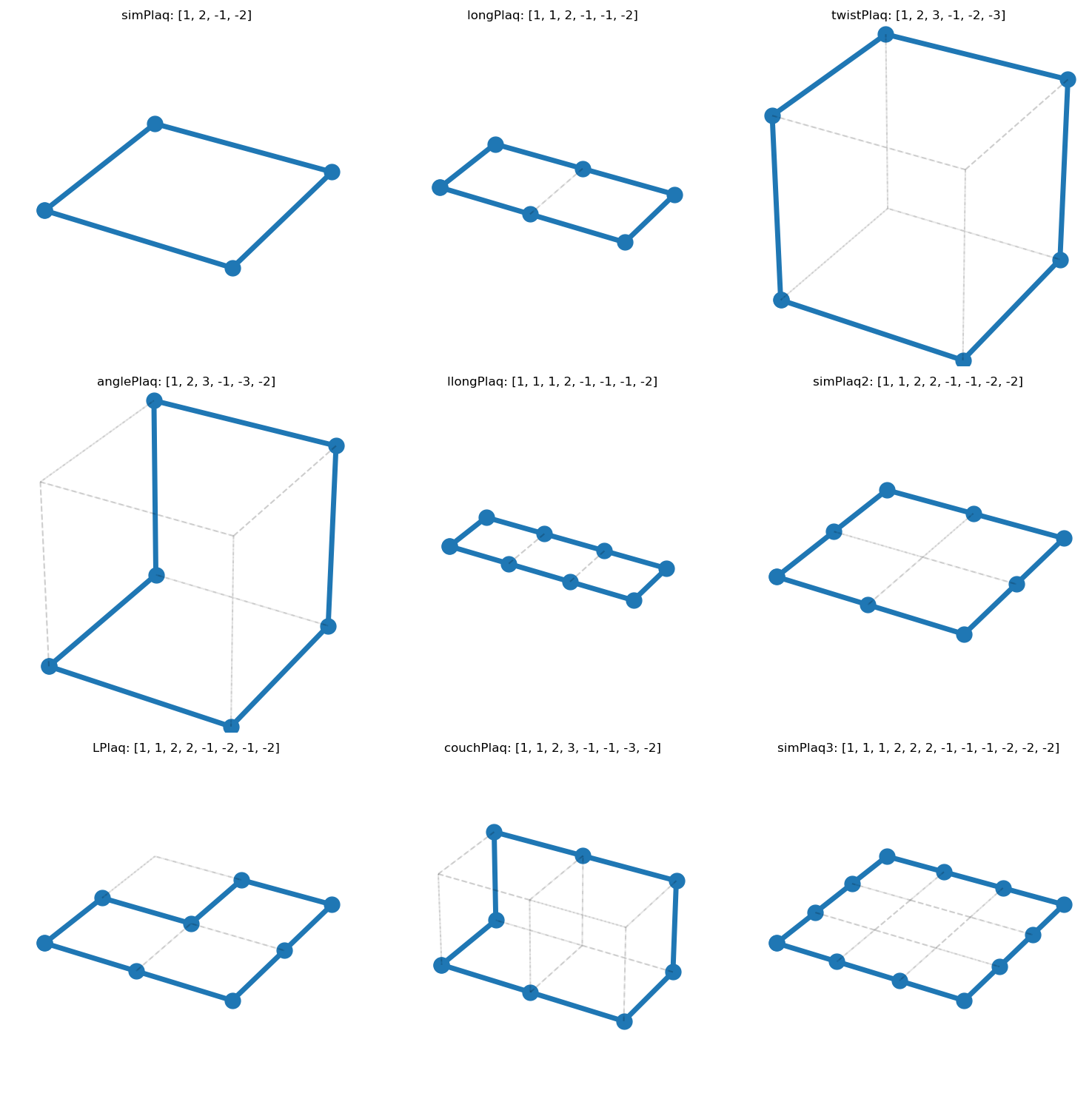}
    \caption{Summary of the loop shapes used in the glueball spectrum measurements.}
    \label{fig:loops}
\end{figure}

We employ a glueball operator basis composed of spatial loops at different levels of (APE) smearing \cite{apeSm}. Measurements are done at $0,2,4,6,8$ smearing levels, with a smearing weight of $\alpha = 1.0$\cite{glue_wf}.
\begin{table}[h]
    \centering
    \begin{tabular}{c c}
        \begin{tabular}{|c|c|c|}
            \hline
            $J^{PC}$ & N. of Operators & after Smearing \\
            \hline
            $A_1^{++}$ & 9 & 45 \\
            \hline
            $A_2^{++}$ & 2 & 10 \\
            \hline
            $E^{++}$ & 20 & 100 \\
            \hline
            $T_2^{++}$ & 12 & 60 \\
            \hline
            $T_1^{+-}$ & 9 & 45 \\
            \hline
            $T_2^{+-}$ & 9 & 45 \\
            \hline
            $A_2^{-+}$ & 2 & 10 \\
            \hline
        \end{tabular}
        &
        \begin{tabular}{|c|c|c|}
            \hline
            $J^{PC}$ & N. of Operators & after Smearing \\
            \hline
            $E^{-+}$ & 2 & 10 \\
            \hline
            $T_1^{-+}$ & 27 & 135 \\
            \hline
            $T_2^{-+}$ & 12 & 60 \\
            \hline
            $A_1^{--}$ & 2 & 10 \\
            \hline
            $E^{--}$ & 4 & 20 \\
            \hline
            $T_1^{--}$ & 3 & 15 \\
            \hline
            $T_2^{--}$ & 9 & 45 \\
            \hline
        \end{tabular}
    \end{tabular}
        \caption{Results of the subduction of the glueball operators onto the cubic group}
        \label{tab:Reduction}
\end{table}
The different loop shapes included in our computation are depicted in figure \ref{fig:loops}. These loops are combined with their rotations to produce linear combinations that transform according to irreducible representations of the symmetry group of the lattice $O_h$. At each lattice site we compute all loops and linear combinations prescribed by group theory, and then sum them over each timeslice to project onto the $p = 0$ mode. These operators form our bases for each representation, the number of operator in each channel is listed in table \ref{tab:Reduction}.

Once we compute the full correlator matrix in each symmetry channel we extract the glueball masses with the usual variational approach. We define the matrix
\begin{align}\label{eq:Corr_Dt}
    C_{ij}(\Delta t) = \left<\left<\mathcal{O}_i(t_0) \mathcal{O}_j(t_0 \pm \Delta t\right>\right>_{t_0, \pm}\,,
\end{align}
where $\left<\cdots\right>_{t_0, \pm}$ indicates a suitable average over the sink position $t_0$ illustrated in section \ref{sec:2lev}.
Then, to extract the spectrum we can look for solutions $v^{(k)}, \lambda^{(k)}$ of the generalized eigenvalue problem (GEVP):
\begin{align}\label{eq:gevp}
    C_{ij}(t_0)v_j^{(k)}(t_0, t_1) = \lambda^{(k)}(t_0, t_1)\cdot  C_{ij}(t_1) v_j^{(k)}(t_0, t_1)\,.
\end{align}

\section{2-level Sampling}\label{sec:2lev}
The locality of the Wilson plaquette action can be exploited to improve the correlator precision in the long time separation region \cite{mltlvl_1,mltlvl_2}.
Indeed the contribution of a single link $U_\mu(x)$ to the Wilson action is:
\begin{equation}
    S(U_\mu(x)) = \frac{\beta}{N}U_\mu(x) V_\mu(x)^\dagger\,,
\end{equation}
where $V_\mu(x)$ is the sum of all the staples that one can build around the link $U_\mu(x)$.\\

\begin{figure}
    \centering
    \includegraphics[width=0.5\linewidth]{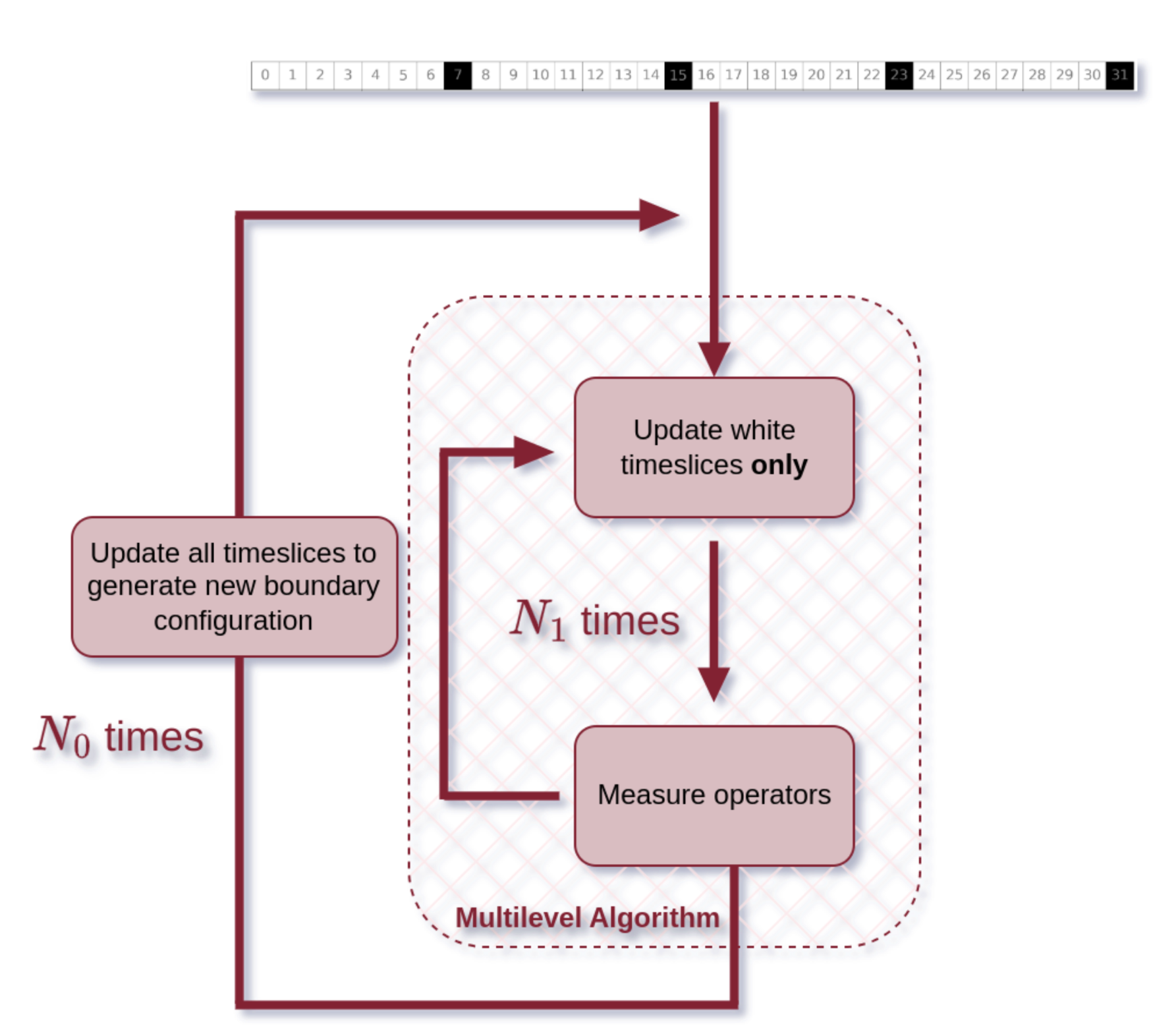}
    \caption{Example flow diagram for 2-level sampling of glueball operators for a lattice of temporal extent of $32a$ and $\Delta = 8$.}
    \label{fig:flowDiag}
\end{figure}

\begin{figure}
    \centering
    \includegraphics[width=0.75\linewidth]{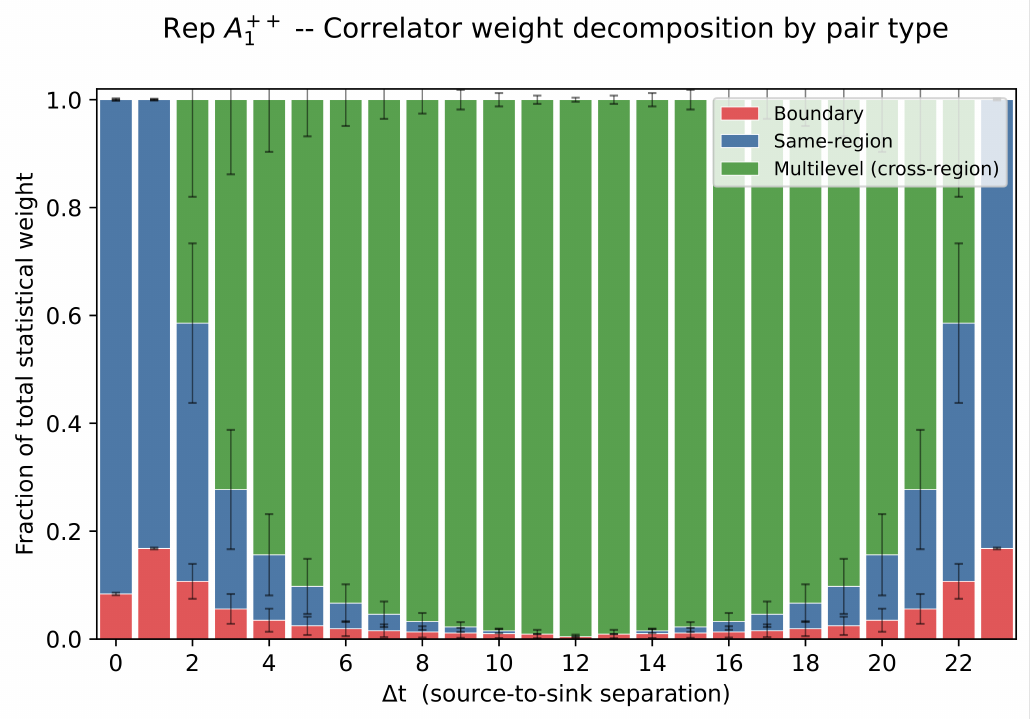}
    \caption{Average distribution of the statistical weights in the sink/source average over diagonal correlators in the scalar channel measured in lattice 2.}
    \label{fig:sinkWeights}
\end{figure}

If we partition our lattice into two distinct regions, $\Lambda_1$ and $\Lambda_2$, separated by fixed boundaries $\partial \Lambda$, in such a way that for every $U_\mu(x)$ the associated $V_\mu(x)$ depends only on link variables located within $\Lambda_1$ or on the boundary\footnote{For the Wilson simple plaquette action, this can be achieved by choosing as the boundary the spatial link variables lying on two different timeslices. For improved actions that include larger Wilson loops, it may be necessary to increase the thickness of the boundary region between the two domains to guarantee their separation, at the cost of having fewer "useful" timeslices to sample.}, and analogously for $\Lambda_2$.
When we are computing the 2-point function between two operators whose timeslices are located in different dynamical regions $\Lambda_i$ this results in the factorization of the path integral
\begin{align}
    \left<\mathcal{O}_i(t_0)\mathcal{O}_j(t_1)\right> &= \int [dU] e^{-S[U]}\mathcal{O}_i(t_0) \cdot \mathcal{O}_j(t_1) = \nonumber\\ &= \int_{\partial \Lambda} [dU]_{\partial \Lambda} e^{-S[U]_{\partial \Lambda}} \left(\int_{\Lambda_1} e^{-S[U]_{\Lambda_1}} \mathcal{O}_i(t_0)\right)\left(\int_{\Lambda_2} e^{-S[U]_{\Lambda_2}} \mathcal{O}_j(t_1)\right)\,.
\end{align}
This means that we can compute the 2-point function as a nested Montecarlo average as depicted in the flow diagram \ref{fig:flowDiag}.
Thus, we produce a set of measurements
\begin{align}
    \left[\mathcal{O}_i^{\mathrm{REP}}(t)_{n_0, n_1}\right]\,,
\end{align}
where REP indicates the symmetry channel in terms of $O_h$ irreducible representations, $i$ is an index running over the whole basis in the REP channel, $n_0 = 1, ..., N_0$ represents the $\partial\Lambda$ configuration and $n_1 = 1, .., N_1$ is the sub-measurement index.\\
When constructing 2-point functions from our measurements we will employ a different strategy depending on the relative sink-source position: if $t_0$ and $t_1$ are in the same dynamical region, or if either of them is on the boundary we will compute the 2-point function in the usual way, considering each of the $N_0\times N_1$ measurements as independent. On the other hand, if $t_0$ and $t_1$ lie in distinct dynamical regions we can compute the 2-point function more accurately with the 2-level nested average:
\begin{equation}
    \label{NestedSumMultilevel}
C^{\mathrm{REP}}_{ij}(t_0,t_1) \equiv \left<\mathcal{O}^{\mathrm{REP}}_i(t) \mathcal{O}^{\mathrm{REP}}_j(t_0)\right> = \frac{1}{N_0 \cdot N_1^2} \cdot \sum_{b = 1}^{N_0} \left(\sum_{\ell = 1}^{N_1} \mathcal{O}_i^{\mathrm{REP}}(t_0)_{b,\ell} \right) \left(\sum_{\kappa = 1}^{N_1}\mathcal{O}_j^{\mathrm{REP}}(t_1)_{b,\kappa} \right)
\end{equation}
Uncertainties on $C^{\mathrm{REP}}_{ij}(t_0,t_1)$ are evaluated using the jackknife method, by subsequently removing from the dataset the data corresponding to one of the $N_0$ boundary configurations.
At this point we average over the sink position, following the procedure described in \cite{mltlvl_2}, to measure the matrix defined in equation \ref{eq:Corr_Dt} by computing the weighted average using $\sigma^{-2}_{C^{\mathrm{REP}}_{ij}(t_0,t_1)}$ as a weight. Correlators computed using nested averaging are naturally less noisy and dominate the long time separation regime of the correlator as shown in figure \ref{fig:sinkWeights} demonstrating the effectiveness of the sampling strategy.

\section{Lattice setup}
We Simulate $SU(6)$ Yang-Mills theory at two different lattice spacings corresponding to $\beta = 25.55, 26.22$ The run parameters and statistics size are listed in table \ref{tab:Lattices}.\\
\begin{table}[h!]
\centering
\begin{tabular}{|c|c|c|c|c|c|c|c|}
\hline
Lat. & $L_0$ & $L_i$ & $\beta$ & $g^2N$ & $N_0$       & $N_1$ & $\Delta_{\mathrm{Multilevel}}$ \\ \hline
1    & 20    & 16    & 25.55   & 2.818  & $\sim$1700  & 100   & 10                             \\ \hline
2    & 24    & 20    & 26.22   & 2.746  & $\sim$ 2000 & 100   & 12                             \\ \hline
\end{tabular}
\caption{Summary table of the lattices and statistics used in the glueball measurements. For meson correlator measurements, a number of configurations have been extracted from the "Lattice 2" ensemble.  }
\label{tab:Lattices}
\end{table}
Montecarlo updates are performed using a composite update sweep, composed of a mixture of Cabibbo-Marinari pseudo-Heatbath and over-relaxation, in a $1:4$ mixture.
The $N_0$ boundary configurations are separated by 128 sweeps, while the $N_1$ configurations obtained through the subupdates are separated by 64 sweeps.\par
All Montecarlo updates use the simple Wilson plaquette action. It is possible to use improved actions in combination with the multilevel algorithm if one chooses the thickness of the frozen timeslices accordingly. Namely, we need to be sure that when we are computing the local contribution to the action from a link in the dynamical region $\Lambda_1$ the computation of the staples $V_\mu(x)$ does not involve any link in $\Lambda_2$.

\subsection{Fermion setup}
In order to compute quark propagators, we invert the Wilson-Dirac lattice operator with periodic boundary conditions
\begin{equation}
D = \left( m + \frac{4}{a} \right) \left( \mathbf{1} - \kappa \sum_{\mu=\pm 1}^{\pm 4} U_{\mu}(n) , \delta_{n+\hat{\mu}, n'} , (\mathbf{1} - \gamma_{\mu}) \right)\,,
\label{eq:DiracWilson}
\end{equation}
where $k$ is the hopping parameter related to the quark mass $m$ via $k=\frac{1}{2(am+4)}$.\\
The quark propagators are estimated stochastically: we utilize four independent noise vectors, $\eta_{i}$, with components drawn from a 
$Z_{2}$ distribution. These sources are localized on the initial time-slice $x_0=0$.
The inversion of the Dirac matrix is performed on a background of gauge configurations previously generated via Monte Carlo simulations with $\beta=26.22$. \\
We employ the Generalized Conjugate Residual (GCR) algorithm, a robust Krylov subspace method. Moreover, to mitigate critical slowing down and accelerate convergence, the GCR solver is preconditioned with the Schwarz Alternating Procedure (SAP) \cite{Luscher_2004}. This domain-decomposition technique partitions the lattice into non-overlapping blocks; the Dirac solution is then updated iteratively through cyclical visits to each subdomain, effectively reducing the condition number of the global operator.\\
All computations are carried out in the quenched approximation. This choice is physically justified in the large-$N$ limit, where the contributions of dynamical fermion loops are suppressed by factors of $\frac{1}{N}$, allowing for high-precision results at a reduced computational cost.

\section{Meson operators}
In this study we are considering a two-flavour non-singlet meson interpolators of the form
\begin{equation}
\ O(n)=\bar{\psi}_{q_{1}}\Gamma\psi_{q_{2}}\,,
\label{eq:GeneralMesonInterpolator}
\end{equation}
where $\Gamma$ is a monomial of Dirac matrices that dictates the quantum numbers of the corresponding meson operator. The symmetry meson channel investigated are the pseudo-scalar and the vector ones:
\begin{table}[h]
    \centering
        \begin{tabular}{|c|c|}
            \hline
            $J^{PC}$ & $\Gamma$ \\
            \hline
            $0^{-+}$ & $\gamma_5$ \\
            \hline
            $1^{--}$ & $\gamma_1$ \\
            \hline
        \end{tabular}
          \caption{Meson symmetry channel and the corresponding Dirac structures used in the computations.}
    \label{tab:MesonChannel}
\end{table}
\\
The 2-point correlator for a meson interpolator on the lattice is equal to
\begin{equation}
G(n_t,m_t) = \sum_{\vec{n}, \vec{m}} \braket{O(n)\bar O(m)}_{F} = -\sum_{\vec{n}, \vec{m}} \mathrm{Tr}\left[\Gamma_{\alpha}S(m,n)_{q_{2}}\bar{\Gamma}_{\beta}S(n,m)_{q_{1}}\right]\
\label{eq:Mesoncorrcomp2}
\end{equation}
where $S(n,m)_{q_{1}}$ and $S(m,n)_{q_{2}}$ are the quark propagators for $\psi_{q_1}$ and $\psi_{q_2}$, respectively, propagating from the source $m$ to the sink $n$. We perform simulations with two light quark flavors with $m_{q_1}=m_{q_2}$.\\
To efficiently numerically evaluate this quantity, we introduce stochastic noise vectors $\eta(m)$ localized at the source time slice $m_t=0$. These vectors satisfy the orthogonality property:
\begin{equation}
\langle \eta_{\alpha,c_{1}}^{\dagger}(m') \eta_{\beta,c_{2}}(m) \rangle_{\text{noise}} = \delta_{m_t',0}\delta_{m_t,0}\delta_{\vec{m}',\vec{m}} \delta_{\alpha,\beta} \delta_{c_{1},c_{2}}
\end{equation}
Using the $\gamma_5$-hermiticity of the Dirac propagator, the analytical expression for the correlator can be rearranged as:
\begin{equation}
G(n_t,m_t) = -\sum_{\vec{n}, \vec{m}} \mathrm{Tr}\left[\Gamma_{\alpha}\gamma_5 S(n,m)_{q_{2}}^{\dagger}\gamma_5\bar\Gamma_{\beta}S(n,m)_{q_{1}}\right]\end{equation}
We then rewrite the trace over the Dirac and color indices using the noise vectors to sum over the spatial source index $\vec{m}$:
\begin{equation}
G(n_t) = G(n_t,m_t=0) =-\sum_{\vec{n}, \vec{m}, \vec{m'} } \langle \eta(m')^{\dagger} \Gamma_{\alpha}\gamma_5 S(n,m')_{q_{2}}^{\dagger}\gamma_5\bar\Gamma_{\beta}S(n,m)_{q_{1}}\eta(m) \rangle_{\text{noise}}
\label{eq:MesonStochSubstitution}
\end{equation}
To compute this efficiently, we define two stochastic inversions propagating from the source time slice $m_t=0$ to the sink $n$:
\begin{equation}
\zeta(n) = \sum_{\vec{m}} S(n,m)_{q_1} \eta(m)\,, \quad \xi(n) = \sum_{\vec{m}} S(n,m)_{q_2} \gamma_5 \Gamma_{\alpha}^{\dagger} \eta(m)
\label{eq:StochasticDirac}
\end{equation}
By averaging over both the noise and gauge configurations, we obtain the final expression for the correlation function $G(n_t)$:
\begin{equation}
G(n_t) = -\sum_{\vec{n}} \left\langle \left(\bar\Gamma_{\beta}^{\dagger}\gamma_5\xi(n)\right)^{\dagger}\zeta(n) \right\rangle_{\! \text{noise, gauge}}
\label{eq:FinalMeson2}
\end{equation}
Consequently, the computation of meson correlators require two Dirac operator inversions per stochastic vector and gauge configuration.
We reduce statistical noise on correlators values by exploiting the time reversal symmetry of the propagators. While gauge ensemble correlations prevent a full doubling of statistics, this symmetry significantly improves precision. 
For a given meson symmetry channel we compute the effective mass by solving the hyperbolic-cosine equation for every $n_t<T/2$:
\begin{equation}
\ \frac{G(n_t)}{G(n_t+1)}=\frac{\cosh(m_{\text{eff}}(n_t-T/2))}{\cosh(m_{\text{eff}}(n_t+1-T/2))}\,.
\label{eq:EffectiveMass}
\end{equation}
This mass varies for different values of $n_t$ and when $G(n_t)$ is dominated by the ground state contribution, $m_{\text{eff}}$ is actually equal to the ground state mass in the given symmetry channel. If we plot in a diagram $m_{\text{eff}}$ as a function of $n_t$, we see a mass effective plateau corresponding to the regime dominated by the ground state.\par
The final ground state mass and its associated error are then determined by performing a constant fit over the identified plateau region. Statistical uncertainties are estimated using the Jackknife resampling method at each time slice.

\begin{table}[h]
    \centering
        \begin{tabular}{|c|c|c|c|}
            \hline
            $J^{PC}$ & N. of Configuration & $k$ \\
            \hline
            $0^{-+}$ & 156 & 0.15479 \\
            \hline
            $1^{--}$ & 155 & 0.15479 \\
            \hline
        \end{tabular}
        \caption{Number of configurations and hopping parameters chosen for the meson simulations}
        \label{tab:Mesonchannelkappa}
\end{table}
\section{Results}
\subsection{Glueballs}
\begin{figure}[htbp]
\centering

\begin{minipage}{0.48\textwidth}
\centering
\begin{tabular}{|c|cc|}
\hline
 &
  \begin{tabular}[c]{@{}l@{}}\textbf{Lattice 1}\\ ($\beta = 25.55$)\end{tabular} &
  \begin{tabular}[c]{@{}l@{}}\textbf{Lattice 2}\\ ($\beta = 26.22$)\end{tabular} \\
                                                  & $a\cdot m$ & $a\cdot m$ \\ \hline
\multicolumn{1}{|c|}{\multirow{2}{*}{$A_1^{++}$}} &   0.68 (4)          & 0.471 (74)  \\
\multicolumn{1}{|c|}{}                            &   /    & 0.9 (4)     \\ \hline
$E^{++}$                                          &    1.13 (19)         & 0.98 (7)    \\ \hline
$T_2^{++}$                                        &     /         & 0.93 (8)    \\ \hline
$T_2^{-+}$                                        &      /       & 0.92 (7)    \\ \hline
\end{tabular}

\captionof{table}{Results of preliminary glueball spectrum extraction in different symmetry channels for both lattices}
\label{tab:resGlue}
\end{minipage}
\hfill
\begin{minipage}{0.48\textwidth}
\centering
\includegraphics[width=\linewidth]{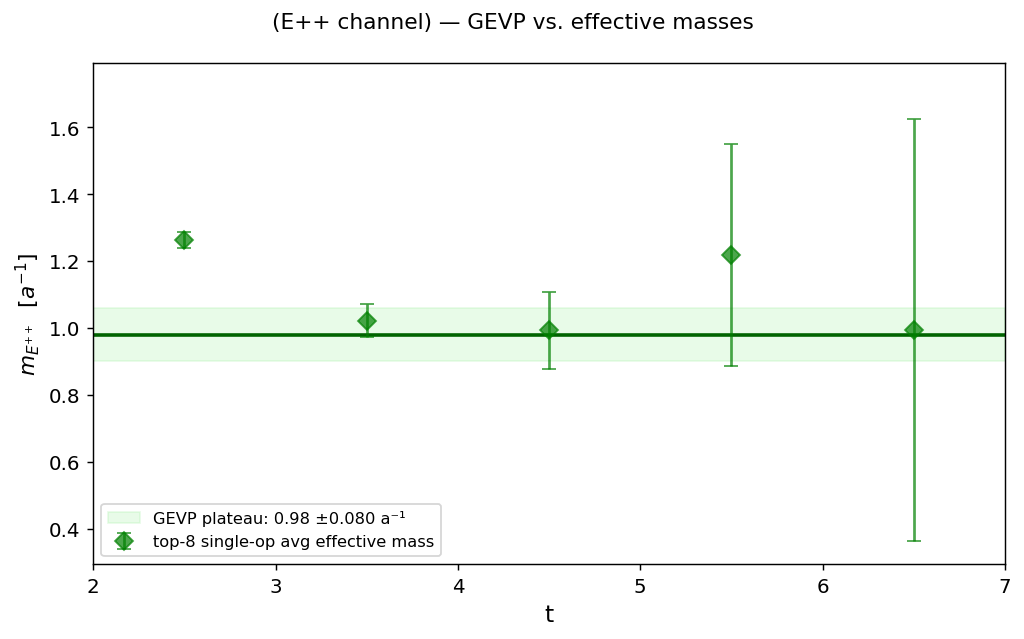}
\captionof{figure}{Example comparison for the $E^{++}$ ground state channel between the GEVP estimate and the effective mass average from the 8 operator with biggest ground state overlap.}
\label{fig:yourLabel}
\end{minipage}

\end{figure}

For symmetry channels where a reliable mass plateau could be identified we list the ground state in table \ref{tab:resGlue}.\\
In almost every channel, spectral analysis of the correlator data at $t_0 =2$ revealed essentially only a single exponential mode surviving the noise, hence the GEVP extraction is carried out on reduced bases, excluding operators with excessive excited state contamination.
\subsection{Mesons}
We report the estimates for the masses of the vector and pseudo-scalar non-singlet mesons. We recall that the final ground state mass and its associated error are then determined by performing a constant fit over the identified plateau region. Statistical uncertainties are estimated using the Jackknife resampling method at each time slice.
\begin{figure}
\centering
\small 

\begin{tabular}{|c|c|}
\hline
\textbf{Channel} & \textbf{$a \cdot m$} \\ \hline
$0^{-+}$  & $0.1869(34)$ \\ \hline
$1^{--}$  & $0.36(9)$ \\ \hline
\end{tabular}
\vspace{-2mm} 
\captionof{table}{Ground state mass results at $\kappa = 0.15479$.}
\label{tab:resMeson}

\vspace{2mm} 

\includegraphics[width=0.35\textwidth]{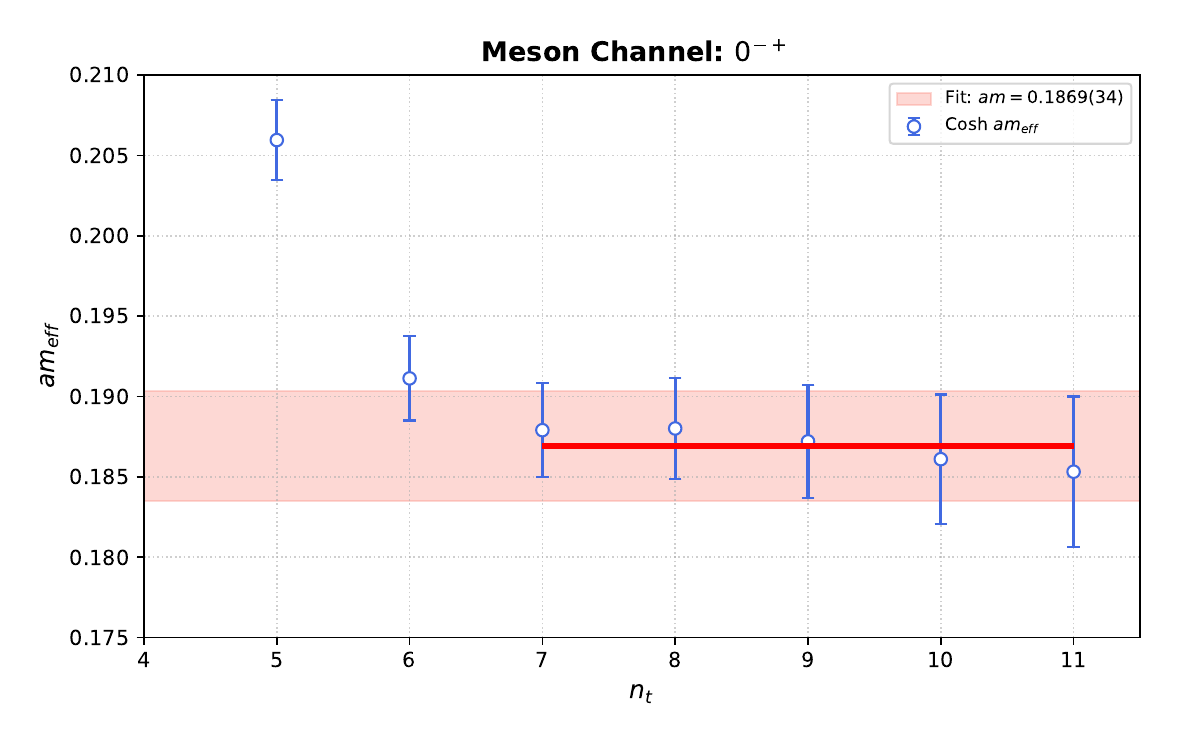}
\hspace{5mm} 
\includegraphics[width=0.35\textwidth]{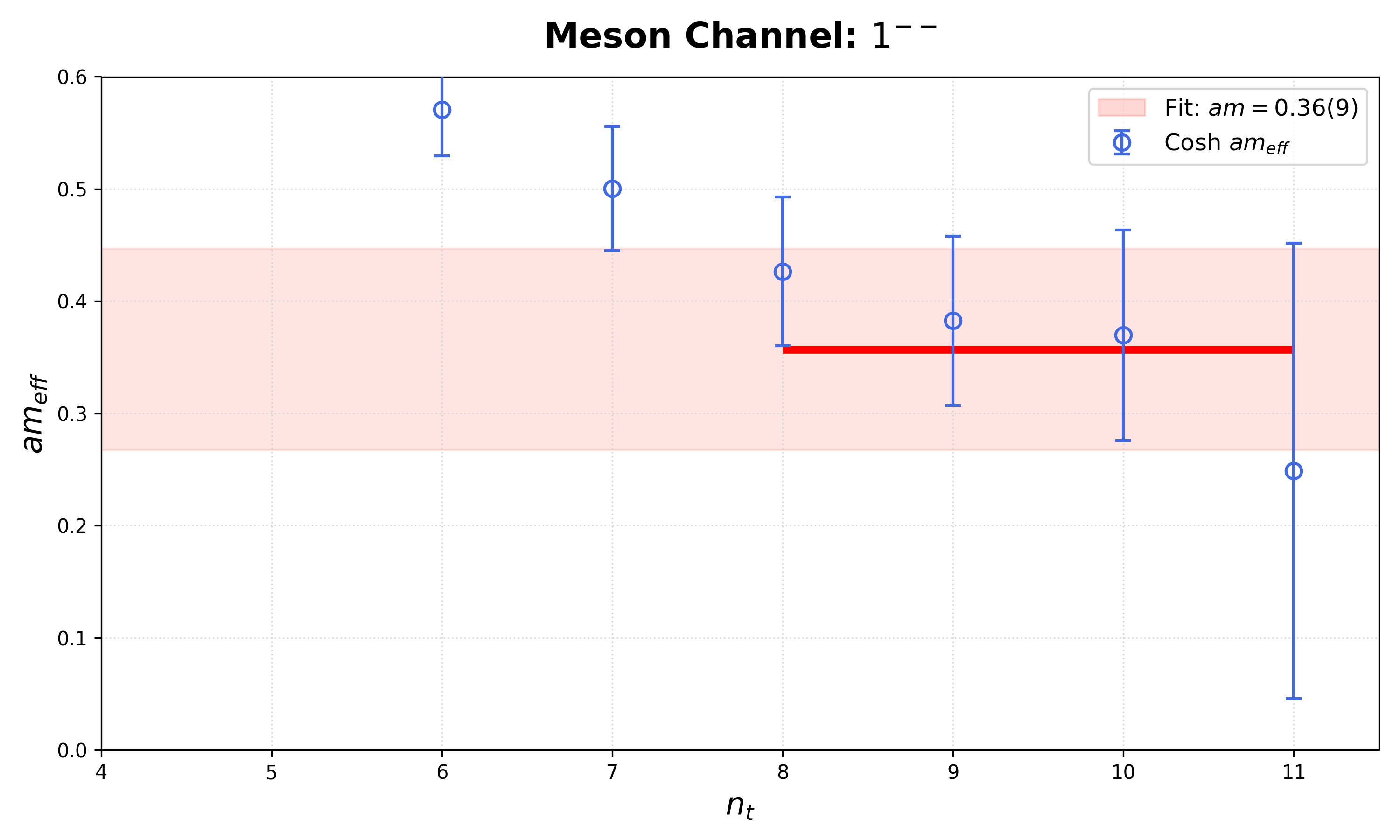}

\vspace{-2mm} 
\captionof{figure}{Effective mass plateaus for $0^{-+}$ (left) and $1^{--}$ (right). }
\label{fig:meff_plots}

\end{figure}

\end{document}